\documentclass[nofootinbib,notitlepage,superscriptaddress,10pt,aps,pra,twocolumn]{revtex4-1}

\usepackage[utf8x]{inputenc}
\usepackage{amsmath}
\usepackage{amssymb}
\usepackage{graphicx}
\usepackage[normalem]{ulem}
\usepackage{epsf,epsfig}
\usepackage{psfrag}
\usepackage{color}
\usepackage{multirow}
\usepackage{diagbox}

\makeatletter
\newcommand\stroke[1]{\mathpalette\stroke@aux{#1}}
\def\stroke@aux#1#2{%
  \ooalign{%
    \hfil$#1^{\;\, \_\hspace{-0.05cm}\_}$\hfil\cr
    \hfil$#1#2$\hfil\cr
  }%
}
\makeatother
\newcommand\dtagliato{\stroke{d}}

\begin{document}

\title{In-vacuo-dispersion features for GRB neutrinos and photons}

\author{Giovanni Amelino-Camelia}
\affiliation{Dipartimento di Fisica, Universit\`a di Roma ``La Sapienza", P.le A. Moro 2, 00185 Roma, Italy}
\affiliation{INFN, Sez.~Roma1, P.le A. Moro 2, 00185 Roma, Italy}
\author{Giacomo D'Amico}
\affiliation{Dipartimento di Fisica, Universit\`a di Roma ``La Sapienza", P.le A. Moro 2, 00185 Roma, Italy}
\affiliation{INFN, Sez.~Roma1, P.le A. Moro 2, 00185 Roma, Italy}
\author{Giacomo Rosati}
\affiliation{Institute for Theoretical Physics, University of Wroc\l{}aw, Pl. Maksa Borna 9, Pl-50-204 Wroc\l{}aw, Poland}
\author{Niccol\'{o} Loret}
\affiliation{Division of Theoretical Physics, Institut Ru\!$\dtagliato$\!er Bo\v{s}kovi\'{c}, Bijeni\v{c}ka cesta 54, 10000 Zagreb, Croatia}

\begin{abstract}
Over the last 15 years there has been considerable interest in the possibility of quantum-gravity-induced
in-vacuo dispersion, the possibility that spacetime itself might behave essentially like a dispersive medium
for particle propagation. Two very recent studies have exposed what might be in-vacuo dispersion features
for GRB (gamma-ray-burst) neutrinos of energy in  the range of 100 TeV and for GRB photons with energy
in the range of 10 GeV. We here show that these two features are roughly compatible with a description such
that the same effects apply over 4 orders of magnitude in energy.
We also characterize quantitatively how rare it would be  for such features
to arise accidentally, as a result of (still unknown) aspects of the  mechanisms  producing photons
at GRBs or as a result of background neutrinos accidentally fitting the profile of
a GRB neutrino affected by in-vacuo dispersion.
\end{abstract}
\maketitle

\section{INTRODUCTION}
The possibility of quantum-gravity-induced in-vacuo dispersion,
an energy dependence of the travel times
of ultrarelativistic particles\footnote{We only consider here photons and high-energy neutrinos,
which are indeed ultrarelativistic particles, particles whose mass is zero or is anyway negligible.}
  from a given source to a given detector,
has been motivated
in several studies
(see {\it e.g.}
Refs.\cite{gacLRR,jacobpiran,gacsmolin,grbgac,gampul,urrutia,gacmaj,myePRL,gacGuettaPiran,steckerliberati} and references therein).
Part of the interest in this possibility comes from the fact that it is a rare example
of candidate quantum-gravity effect that could lead to observably large manifestations, even if,
as it appears to be safe to assume, its characteristic length scale is of the order of the
minute Planck length (inverse of the Planck energy scale $E_P \simeq 10^{28}eV$) or anyway not much larger than that.

The best opportunity so far studied for such experimental tests is provided by observations
of GRBs \cite{gacLRR,jacobpiran,gacsmolin,grbgac}, which set up for us a sort of race among
photons of different energies and (probably \cite{waxbig,meszabig,dafnebig,otherbig}) neutrinos
of different energies, all emitted within a relatively small time window. The fact that our understanding
of the mechanisms producing GRBs remains preliminary is a challenge, since any given time-of-arrival difference
 among two particles can in  principle always be attributed to the emission mechanism,
 but this
can be compensated by suitable techniques of statistical analysis.

For more than a decade the analyses of GRB data from the in-vacuo-dispersion perspective
were done considering only photons and focusing on what could be tentatively
inferred from each single GRB. Recently, thanks mainly to the IceCube telescope, it became possible
to contemplate the possibility that we might be observing also some GRB neutrinos affected
by in-vacuo dispersion; moreover, for GRB photons the abundance of observations cumulatively obtained
by the Fermi telescope reached the level sufficient for attempting to perform statistical
analyses over the whole collection of Fermi-observed GRBs.
Some of us were involved in the first studies
using IceCube data for searching for GRB-neutrino in-vacuo-dispersion
candidates~\cite{gacGuettaPiran,Ryan,RyanLensing}.
Intriguing  statistical
analyses of in-vacuo dispersion over the whole collection of Fermi-observed GRBs
were perfomed in a series of studies by Bo-Qiang Ma and collaborators~\cite{MaZhang,MaXuPRIMO,MaXuSECONDO}.
The neutrino studies of Refs.~\cite{gacGuettaPiran,Ryan,RyanLensing} led to exposing a feature
in the IceCube neutrino data which could plausibly be a manifestation of in-vacuo dispersion.
The possibility that this feature could be the result of background neutrinos just accidentally
arranging themselves as if they were GRB neutrinos affected by in vacuo dispersion was considered
using statistical tools of analysis, finding that it would be ``very untypical" for background
neutrinos to produce accidentally such a pronounced in-vacuo dispersion feature.
The GRB-photon studies reported in Refs.~\cite{MaZhang,MaXuPRIMO,MaXuSECONDO}
also led to exposing a feature which could be a manifestation of in-vacuo dispersion.
While this feature for GRB photons is certainly striking, as observed most convincingly
in  Ref.~\cite{MaXuSECONDO}, there was so far no attempt
to characterize quantitatively its statistical significance.

The main objective of the study we are here reporting is to characterize the
statistical significance of the feature exposed in Refs.~\cite{MaZhang,MaXuPRIMO,MaXuSECONDO}
for photons,
and to show that this feature is surprisingly consistent with the feature exposed
in Refs.~\cite{gacGuettaPiran,Ryan,RyanLensing} for neutrinos of much higher energies (the relevant photons
have energies of the order of 10 GeV, while the neutrinos have energies
of the order of 100 TeV).
We also offer some preliminary observations which might become relevant if any of the features
here contemplated find greater support as more data are accrued, concerning the possible
interpretation of such features as manifestations of (so far unknown) astrophysical mechanisms,
rather than as manifestations of in-vacuo dispersion.

\section{Modeling quantum-gravity-induced in-vacuo dispersion}
The class of scenarios we intend to contemplate finds motivation in
some much-studied models of
spacetime quantization (see, {\it e.g.}, \cite{jacobpiran,gacsmolin,gacLRR,grbgac,gampul,urrutia,gacmaj,myePRL} and references therein)
 and, for the type of data analyses we are interested in, has the implication that
 the time needed for a ultrarelativistic particle
to travel from a given source to a given detector receives a quantum-spacetime correction, here denoted with $\Delta t$.
We focus on the class of scenarios whose predictions for energy ($E$) dependence of $\Delta t$ can all be described
in terms of the formula
(working in units with the speed-of-light scale ``$c$" set to 1)
\begin{equation}
\Delta t = \eta_X \frac{E}{M_{P}} D(z) \pm \delta_X \frac{E}{M_{P}} D(z) \, .
\label{main}
\end{equation}
Here the redshift- ($z$-)dependent  $D(z)$ carries the information on the distance between source and detector, and it factors
in the interplay between quantum-spacetime effects and the curvature of spacetime.
As usually done in the relevant literature \cite{jacobpiran,gacsmolin,gacLRR} we take for $D(z)$ the following form:\footnote{The interplay between quantum-spacetime effects and curvature of spacetime is still a lively subject of investigation, and, while (\ref{dz})
is by far the most studied scenario, some alternatives to (\ref{dz}) are also under consideration \cite{dsrfrw}.}
\begin{equation}
D(z) = \int_0^z d\zeta \frac{(1+\zeta)}{H_0\sqrt{\Omega_\Lambda + (1+\zeta)^3 \Omega_m}}  \, ,
\label{dz}
\end{equation}
where $\Omega_\Lambda$, $H_0$ and $\Omega_0$ denote, as usual,
respectively the cosmological constant, the Hubble parameter and the matter fraction, for which we take the values given in Ref.\cite{PlanckCosmPar}.
With $M_{P}$ we denote the Planck scale ($\simeq 1.2\,\cdotp 10^{28}eV$) while
the values of the  parameters $\eta_X$ and $\delta_X$ in (\ref{main})
characterize the specific scenario one intends to study. In particular, in (\ref{main}) we used the notation ``$\pm \delta_X$"
to reflect the fact that $\delta_X$ parametrizes the size of quantum-uncertainty (fuzziness) effects. Instead the parameter $\eta_X$
characterizes systematic effects: for example in our conventions for positive $\eta_X$ and $\delta_X =0$ a high-energy particle
is detected systematically after a low-energy particle (if the two particles are emitted simultaneously).

The dimensionless parameters $\eta_X$ and $\delta_X$ can take different values for different types of particles \cite{gacLRR,myePRL,mattiLRR,szabo1}, and it is of particular interest for our study that in particular for
neutrinos some arguments have led to the expectation of an helicity dependence of the effects (see, {\it e.g.},
Refs.\cite{gacLRR,mattiLRR} and references therein). Therefore even when focusing only on neutrinos one should
contemplate four parameters, $\eta_+$, $\delta_+$, $\eta_-$, $\delta_-$ (with the indices $+$ and $-$ referring of course
to the helicity). Analogous considerations apply to photons and their polarization \cite{gacLRR,mattiLRR}.
The parameters $\eta_X,\delta_X$ are to be determined experimentally. When non-vanishing,
they are expected to take values somewhere in a neighborhood of 1, but values as large as $10^3$ are plausible if the solution
to the quantum-gravity problem is somehow connected, as some arguments suggest \cite{gacLRR,wilczek,hsuHIGGSES},
with the unification of non-gravitational forces,
while values
 smaller than 1 find support in some renormalization-group arguments (see, {\it e.g.}, Ref.\cite{hsuHIGGSEStwo}).

Following Refs.~\cite{MaZhang,MaXuPRIMO,MaXuSECONDO}, we
 find convenient to introduce a ``distance-rescaled energy" $E^*$ defined as\footnote{While here and
 in Refs.~\cite{MaZhang,MaXuPRIMO,MaXuSECONDO} the analysis is set up in terms
 of correlations between $\Delta t$ and a ``distance-rescaled energy" $E^*$, in
 Refs.~\cite{gacGuettaPiran,Ryan,RyanLensing}, which focused on neutrinos, the analysis was set up in terms
 of correlations between energy and a ``distance-rescaled time delay" $\Delta t^*$. The two setups are evidently equivalent,
 but the one we adopt here is best suited for handling the possibility of a (roughly-)systematic time offset
 at the source (see later). For the values of $\Delta t$ that are relevant for the neutrino part
 of the analysis this possibility of a time offset has a negligible role, and therefore the two setups are actually
 equally convenient, but for part of the analysis based on photons it is advantageous to set up
 the analysis in terms of correlations between $\Delta t$ and a ``distance-rescaled energy" $E^*$.}
\begin{equation}
E^* \equiv E \frac{D(z)}{D(1)}
\label{tstar}
\end{equation}
so that (\ref{main}) can be rewritten as
\begin{equation}
\Delta t = \eta_X D(1) \frac{E^*}{M_{P}}  \pm \delta_X D(1) \frac{E^*}{M_{P}}  \, .
\label{maintwo}
\end{equation}
This reformulation of (\ref{main}) allows to describe the relevant quantum-spacetime effects,
which in general depend both on redshift and energy, as effects that depend exclusively on energy,
through the simple expedient of focusing on the relationship between $\Delta t$ and energy
when the redshift has a certain chosen value, which in particular we chose to be $z=1$.
If one measures a certain $\Delta t$ and the redshift $z$ of
the relevant GRB is well known, then one gets a firm determination of $E^*$ by simply rescaling
the measured $E$ by the factor $D(z)/D(1)$. And even when the redshift
of the relevant GRB is not known accurately one will be able to convert a measured $E$
into a determined $E^*$ with accuracy governed by how much one is able to still assume
about the redshift of the relevant GRB. In particular, even just the information on whether a GRB is long
or short can be converted into at least a very rough estimate of redshift.

Eq.(\ref{maintwo}), which follows the strategy of analysis proposed in Refs.~\cite{MaZhang,MaXuPRIMO,MaXuSECONDO},
is ideally structured to handle the possibility that there be a (roughly) systematic time offset at
emission between the time of emission of
the low-energy particles used as reference (we shall later take as reference the time of
observation of the first peak of  the  low-energy-gamma-ray component of a GRB)
and the higher-energy particle of interest. Such an astrophysical mechanism for time offset at the source, would
imply, within the modelization we are assuming for the quantum-spacetime effects,
that $\Delta t$ is not exactly proportional to $E^*$, since the observed $\Delta t$ would receive both a contribution
 from the quantum-spacetime effects given by the right-hand side of Eq.(\ref{maintwo}) and a contribution
 due to the time offset at the source. This latter contribution can be described as $(1+z) t_{off}$, where
 $t_{off}$ is the time offset at the source and the factor $(1+z)$ takes into account
 time dilatation. Following Refs.~\cite{MaZhang,MaXuPRIMO,MaXuSECONDO}
 these observations can be fruitfully used to replace Eq.(\ref{maintwo}) with
\begin{equation}
\frac{\Delta t}{1+z}=t_{off}+ \frac{\eta_X}{M_{P}} D(1) \frac{E^*}{(1+z)}  \pm \frac{\delta_X}{M_{P}} D(1) \frac{E^*}{(1+z)} \, .
\label{mainnew}
\end{equation}
Notice that in allowing for the mentioned possibility of a time offset at the source we also found appropriate
to set up our equation as a relationship between $\frac{\Delta t}{1+z}$ and $\frac{E^*}{(1+z)}$, so that the
term involving $t_{off}$ is just a constant contribution, redshift independent and energy independent.
Later, in our graphs showing $\frac{\Delta t}{1+z}$ versus $\frac{E^*}{(1+z)}$, this will facilitate
the visualization of $t_{off}$. We stress that here, just like in Refs.~\cite{MaZhang,MaXuPRIMO,MaXuSECONDO},
we shall not allow for different values of $t_{off}$ for different photons\footnote{The interested  reader
can easily see that by allowing different values of $t_{off}$ for different photons one could never
test the in-vacuo-dispersion hypothesis, since any measured value of $\Delta t$ could always be attributed
to a corresponding value of $t_{off}$ at the source.}. We just allow for one  value of $t_{off}$
valid for all photons of all GRBs in the analysis, and we shall show
that the present data situation fits rather nicely this apparently simplistic assumption.

\section{Summary of previous analysis of GRB-neutrino candidates}
As stressed above, the main objective of the study we are here reporting is to characterize the
statistical significance of the in-vacuo-dispersion
feature exposed in Refs.~\cite{MaZhang,MaXuPRIMO,MaXuSECONDO}
for photons,
and to show that this feature is surprisingly consistent with the feature exposed
in Refs.~\cite{gacGuettaPiran,Ryan,RyanLensing} for neutrinos.
A quantitative characterization of the statistical significance of the in-vacuo-dispersion
feature found for neutrinos was already given in Ref.~\cite{Ryan}, so we shall not have new  results about that
here, but it is still valuable for our purposes to summarize the main steps of that analysis.
This will also give us a chance to arrange the presentation in terms of correlations
between $\Delta t/(1+z)$ and $E^*/(1+z)$, whereas
in Ref.~\cite{Ryan} the analysis was arranged in terms of correlations between energy and
a ``distance-rescaled time-of-arrival difference". These two arrangements of the analysis
are evidently equivalent for the neutrino case\footnote{The two arrangements of the analysis
are completely equivalent for our neutrinos, since for them the hypothesis of a time offset at the
source is irrelevant, as we shall soon observe. For photons, were a time offset at the source
could have tangible consequences, it is truly convenient
(as first observed in Refs.~\cite{MaZhang,MaXuPRIMO,MaXuSECONDO})
 to arrange the analysis in terms of
correlations between $\Delta t/(1+z)$ and $E^*/(1+z)$.},
but it is a good preparation for the later discussion of the photon case to
have the discussion of neutrinos arranged in terms of correlations between $\Delta t/(1+z)$ and $E^*/(1+z)$.

For the neutrino case a crucial role is played by the criteria used for selecting some GRB-neutrino candidates.
This is not at all an easy task since the present situation is such that at best we can catch a single neutrino
from a whole GRB. Moreover, in testing the hypothesis of in-vacuo dispersion, we must allow for a sizable
time-of-observation difference between the neutrino and the first peak in Fermi's GBM\footnote{The lowest-energy part
of Fermi's observations are performed by the GBM. We follow Refs.~\cite{MaZhang,MaXuPRIMO,MaXuSECONDO}
is adopting the first peak of the GBM as the reference time of observation of a GRB, since we feel it is a rather
natural criterion, already adopted by other authors in previous studies, which we found no
good reason to modify. We shall however stress that this criterion plays basically no role
for our neutrino part of the discussion and plays only a rather small role for the photon part.}. In fact,
at the scales of interest for this neutrino analysis, involving
in particular neutrinos with energy of order 100 TeV,
in-vacuo dispersion could produce values of $\Delta t$ of anything between a few hours and a couple
of days.

Of course  such  criteria for selecting GRB-neutrino candidates will  involve
a temporal window (how large can the $\Delta t$ be in order for us to consider a
IceCube event as a potential GRB-neutrino candidate) and some criteria of directional selection (how well the directions estimated
for the IceCube event and for the GRB should agree in order for us to consider that IceCube event as a potential GRB-neutrino candidate).
A previous more preliminary analysis (based on IceCube data from June 2008 to May 2010)
had tentatively put in focus a range of values of $\eta_\nu$ (for our Eq.(\ref{main})
somewhere in the range between 10 and 20, and this we used in
Refs.~{\cite{Ryan,RyanLensing}} (based
on IceCube data from June 2010 to May 2014) to choose a temporal window large enough
to accommodate the corresponding size of the effects.
We took a temporal window of 3 days, and  focused on IceCube events with energy
between 60 TeV\footnote{The 60-TeV lower limit of our range of energies is consistent
with the analogous choice made by other studies
whose scopes, like ours, require keeping the contribution of background neutrinos relatively low \cite{IceCube,IceCubeBackground}.} and 500 TeV.
Widening the range of energies up to, say, 1000 TeV would have imposed us a temporal window
of about 6 days, rendering even more severe one of the key challenges for this sort of analysis, which is the one of multiple
GRB candidate partners for a single IceCube event. As directional criteria for the selection of GRB-neutrino candidates we considered the signal direction PDF depending on the space angle difference between GRB and neutrino: $P(\nu,GRB)=(2\pi\sigma^2)^{-1}\exp(-\frac{|\vec{x}_{\nu}-\vec{x}_{GRB}|^2}{2\sigma^2})$, a two dimensional circular Gaussian whose standard deviation is  $\sigma=\sqrt{\sigma_{GRB}^2+\sigma_{\nu}^2}$, asking
 the pair composed by the neutrino and the GRB to be at angular distance compatible within a 2$\sigma$ region.

A key observation for our analysis (based on the corresponding observation made in our
Ref.~\cite{Ryan}) is that whenever $\eta_X$ and/or $\delta_X$
 do not vanish one should expect
on the basis of (\ref{mainnew}) a correlation between $|\Delta t|/(1+z)$ and $E^*/(1+z)$.

Our data set~\cite{Ryan} is for four years of operation of IceCube \cite{IceCube},
  from June 2010 to May 2014.
Since the determination of the energy of the neutrino plays such a  crucial role in our analysis
we include only IceCube ``shower events" (for ``track events" the reconstruction of the neutrino energy is far more problematic  and less reliable \cite{IceCubeBackground,TRACKnogood1}).
We have 21 such events within our 60-500 TeV energy window, and we find that 9 of them
fit our requirements for candidate GRB neutrinos.
The properties of these 9 candidates that are most relevant for our analysis are summarized in Table 1 and Figure 1.

\begin{table}[htbp]
\centering
{\def\arraystretch{0.5}\tabcolsep=3pt
\begin{tabular}{|c|c|c|r|l|c|c|}
\hline
$\,$                   &    \!\!\!   E \!\!\!\! [TeV]      \!\!\!  &     \!\!\!   $E^*$ \!\!\!\! [TeV]     &    $  \Delta t$ [s]        & z           &   GRB & $\,$   \\\hline
IC9                          &                 63.2       &  101.1          &  80335    & 1.613        &    110503A   &     *    \\\hline
IC19                         &                71.5        & 98.5          & 73960   & 1.3805        &     111229A     &     *  \\\hline
\multirow{3}{*}{IC42} & \multirow{3}{*}{76.3}    &   273.3  &  20134     & 4.042    &   131117A     & $\,$     \\
&                                       									& 113.6  &   -146960    & 1.497 *   &    131118A      &     *    \\
&                                       									& ? &   -218109    & \,\,\,\,\, ? &   131119A          &   $\,$      \\
\hline
IC11                          &                88.4        & 131.7         &  185146   & 1.497   *     &   110531A     &     *     \\\hline
IC12                         &                104.1       & 155.0          &  160909 & 1.497   *     &    110625B     &     *    \\\hline
\multirow{3}{*}{IC2} & \multirow{3}{*}{117.0} & ?    &  15445    & \,\,\,\,\, ?  &     100604A        & $\,$     \\
                                &                                     & 174.2   &   -113051    & 1.497 *    &     100605A   &     * \\
                                &                                     & ?   &  -201702    & \,\,\,\,\, ?  &    100606A    & $\,$           \\\hline

IC40                          &             157.3         & 234.3        &   -179641  & 1.497   *     &    130730A    &     *    \\\hline
\multirow{2}{*}{IC26}
& \multirow{2}{*}{210.0}  & 312.8 &   229039     & 1.497  *  &  120219A   &     *    \\
&                        & ?  &  -175141    & \,\,\,\,\, ? &   120224B    &  $\,$  \\
\hline
IC33                          &             384.7       & 227.4         &  -171072   & \,\, 0.6      *     &   121023A      &     *   \\\hline
\end{tabular}
}
\caption{Among the 21 ``shower neutrinos" with energy between 60 and 500 TeV observed by IceCube between June 2010 and May 2014
only 9 fit our directional and temporal criteria for GRB-neutrino candidates. For 3 of them there is more than one GRB
to be considered when pairing up neutrinos and GRBs. The last column highlights with an asterisk the 9 GRB-neutrino candidates ultimately selected in our Ref.\cite{Ryan}
by our additional criterion of maximal correlation.
Also shown in table are, when known, the values of redshift attributed to the relevant GRBs.
The only redshift measurements relevant for our 9 GRB-neutrino candidates are those
for GRB111229A and GRB110503A, which are long GRBs, and we assume that the average of their redshifts (1.497) could be a reasonably good estimate
of the redshifts of the other long GRBs relevant for our 9 GRB-neutrino candidates. These are the 6 estimated values of redshift
$z=1.497^*$, the asterisk reminding that it is a ``best guess" value. For analogous reasons we place an asterisk close to the value of 0.6 which is our best guess for the redshift of the only short GRB
in our sample. The first column lists the ``names" given by IceCube to the relevant neutrinos,
with their observed energies reported in the second column.
Third and fourth column give the values of the $\Delta t$ and $E^*$ defined  in the main text.}
\label{table1}
\end{table}

\begin{figure}[h!]

\includegraphics[scale=0.6]{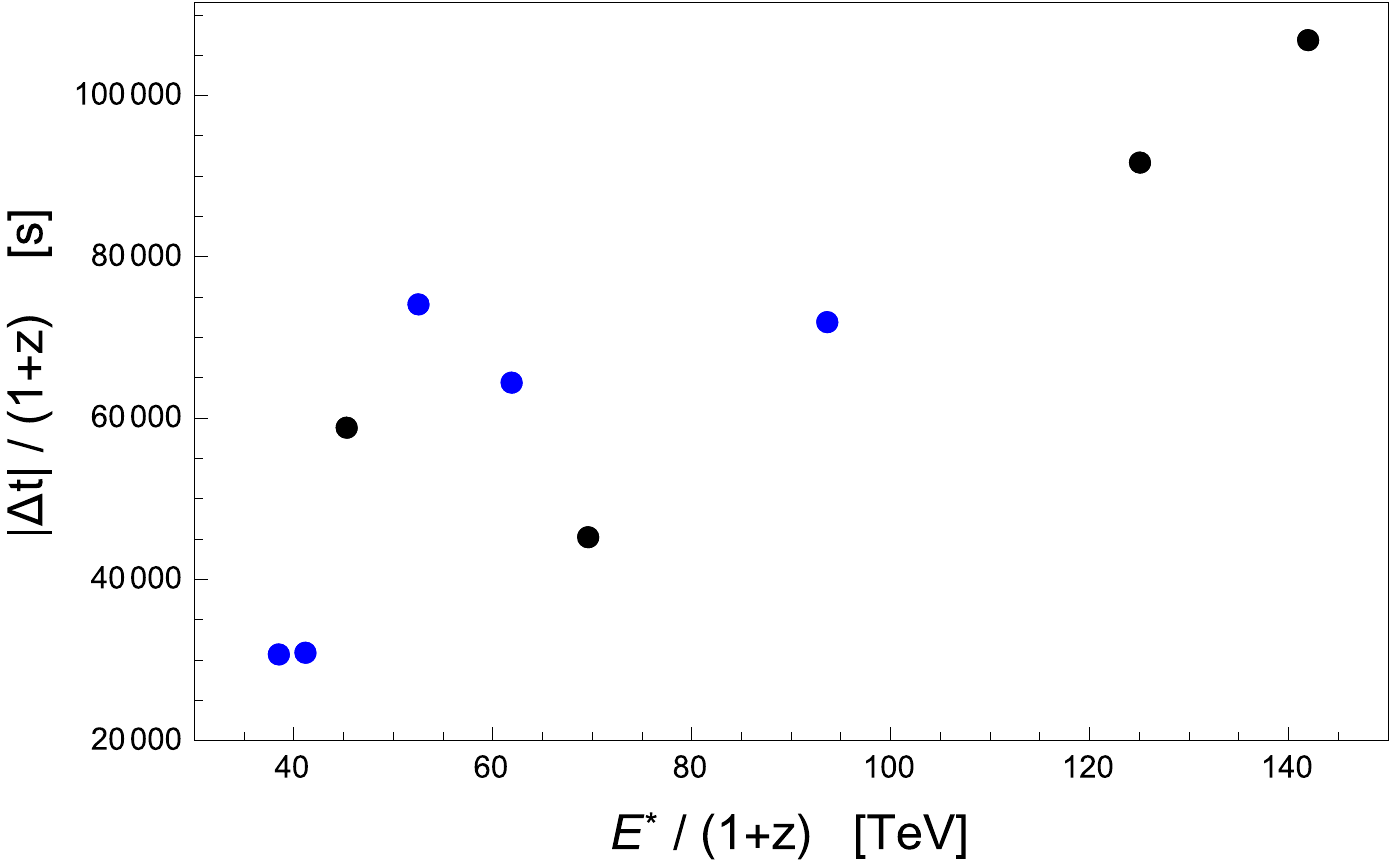}
\caption{Points here in figure correspond to the 9 GRB-neutrino candidates highlighted with an asterisk in the last column of Table 1. Blue points correspond to ``late neutrinos" ($\Delta t>0$),
while black points correspond to ``early neutrinos" ($\Delta t<0$).}
\end{figure}

As visible in Table 1,
for some IceCube events our selection
criteria produce multiple GRB-neutrino candidates.
In Ref.\cite{Ryan} we handled this issue of multiple candidates
by focusing on the case
 that provides the highest correlation.

Another issue reflected by Table 1 comes from the fact that for only 3 of the GRBs involved in this analysis the redshift is known.  We must handle only one short GRB of unknown redshift, and we assume for it a redshift of 0.6,
which is a rather reasonable rough estimate for a short GRB. Our 9 GRB-neutrino candidates marked by an asterisk in table 1
 include 8 long GRBs, 2 of which have known redshift, and we assign to the other 6 long GRBs
the average ${\bar{z}}$  of those two values of redshift (${\bar{z}}=1.497$).

In figure 1 it is striking that the correlation  between $|\Delta t|/(1+z)$ and $E^*/(1+z)$
gets stronger at higher energies. Interestingly, as observed in our Ref.\cite{Ryan}, this too fits
the expectations of some quantum-spacetime models: as stressed in particular in Ref.\cite{steckerliberati},
in some of these quantum-spacetime models neutrinos can undergo processes of ``neutrino splitting",
and in turn this could plausibly \cite{Ryan} affect a in-vacuo-dispersion study such as ours just in
the way of rendering the correlation weaker at lower energies. While this was worth mentioning, we shall
here prudently not take it into account: we shall ignore neutrino splitting and handle all our 9
GRB-neutrino candidates on the same footing.

The correlation between $|\Delta t|/(1+z)$ and $E^*/(1+z)$
 for the 9 GRB-neutrino candidates highlighted in Fig.1
 is of\footnote{ In Ref.\cite{Ryan}, where the correlation study was arranged for energy versus a time-of-observation difference rescaled by a function of redshift, we had found for the same 9 candidates a correlation of 0.951.
 This 0.951 goes down to 0.866 when arranging the analysis for correlation between $\Delta t/(1+z)$ and $E^*/(1+z)$.
 These two types of correlation studies are based on two equations which are equivalent to each other,
 one  obtained from the other by simply dividing both members of the equation by the same function of redshift.
Therefore in the ideal case of an infinite amount of data the indications emerging from the two types of correlation
studies would be exactly coincident, but only 9 data points intervene in  our analysis, spread over a wide range
of values of redshift, and this results in the (however small) difference between 0.951 and 0.866.} 0.866.
This is a strikingly high value of correlation but in itself
does not provide what is evidently the most interesting quantity here of interest,
which must be some sort of ``false alarm probability": how likely it would be to have accidentally data with such
good agreement with the expectations of the quantum-spacetime models here contemplated?
We proposed in Ref.\cite{Ryan} that for these  purposes one could estimate how often
a sample composed exclusively of background neutrinos\footnote{Consistently with the objectives of our analysis we consider as ``background neutrinos" all
 neutrinos that are unrelated to a GRB, neutrinos of atmospheric or other astrophysical origin which end up being selected
as GRB-neutrino candidates just because accidentally their time of detection and direction of detection happen to fit
our selection criteria.} would produce accidentally
9 or more GRB-neutrino candidates with correlation
comparable to (or greater than) those we found in data.
We did this by performing $10^5$ randomizations of the times of detection
of the 21 IceCube neutrinos relevant for our analysis, keeping their energies and directions fixed,
and for each of these time randomizations we redo the analysis\footnote{In particular for any given realization of the fictitious GRB-neutrino candidates we identify those of known redshift and use them to estimate the ``typical fictitious GRB-neutrino redshift", then attributed to
those candidates of unknown redshift (procedure done separately for long and for short GRBs). When in the given
  realization of the fictitious GRB-neutrino candidates there is no long (short) GRB of known redshift we attribute to all of them
a redshift of 1.497 (0.6).} just as if they were real data.
Our observable is a time-energy correlation and by randomizing the
times we get a robust estimate of how easy (or how hard) it is for a sample composed exclusively
of background neutrinos to produce accidentally
a certain correlation result.
Also in the analysis of these fictitious data obtained by randomizing the detection times of the neutrinos
we handle cases with neutrinos for which there is more than one possible GRB partner by maximizing the correlation,
in the sense already discussed above for the true data. We ask how often this time-randomization
procedure produces 9 or more GRB-neutrino candidates with correlation
$\geq 0.866$, and remarkably we find that this happens only in 0.11$\%$ of cases.

Having correlation as high as 0.866 (and false alarm probability of 0.11$\%$) is particularly striking
considering that surely at least some of our 9 GRB-neutrino candidates are just background neutrinos
accidentally fitting our criteria for the selection of GRB-neutrino candidates.
This can be straightforwardly deduced by observing that out of the 21 neutrinos in our sample, since indeed
only 9 turned out to fit
our requirements for GRB-neutrino
candidates, there are at least 12 neutrinos which are background\footnote{Importantly those 12 neutrinos are background
in both pictures here relevant: if the model of Eq.(\ref{main}) is correct they are to be considered as background since they
do not fit our selection criteria, and of course those 12 neutrinos are background also if Eq.(\ref{main}) is not correct
 (in that case all the 21 events are not GRB neutrinos).}.
We can therefore ask how likely it would have been for one or more of those 12 neutrinos to accidentally appear
to be GRB neutrinos of the type we are looking for. This can be estimated by randomizing the times of those 12 neutrinos.
Of course, if, say, it is likely that 3 of those neutrinos could appear as GRB neutrinos, we will assume
that a proportionate number of our 9 GRB-neutrino candidates are background.
Details on this line of reasoning are  given in  Ref.~\cite{RyanLensing}. Important for us here is that this line of reasoning
leads to the conclusion that it is very likely that at least 3 of our 9 GRB-neutrino candidates are background.
This renders somewhat striking the fact that, in spite of these contributions by background neutrinos,
we found a value of correlation as high as 0.866.

Concerning our notion of ``false alarm probability" this deduction about the role played by background
neutrinos may suggest a somewhat different strategy of analysis~\cite{RyanLensing}. Let us illustrate this by taking as
working assumption that 3 among our 9 GRB-neutrino candidates surely are background.
We can then exploratively assume that the 6-plet of ``true" GRB neutrinos is the maximum-correlation 6-plet
among the 6-plets obtainable from our 9 candidates, and take as reference for the
analysis the value of correlation found for this maximum-correlation 6-plet, which is 0.995.
One can then define a false alarm probability based on how frequently simulated data, obtained by randomizing the times
of detection of all the 21 neutrinos in our sample, include a 6-plet of candidates with correlation greater or equal to
the value of 0.995 found for the best 6-plet in the real data
(so, if, say, a given time randomization produces 11 candidates one
would assign to the randomization a value of correlation given by the highest correlation found by considering all possible choices
 of 6 out of the 11
candidates). We find that this false alarm probability is of 0.6$\%$.

\bigskip

\section{In-vacuo dispersion for high-energy Fermi-telescope photons}

\subsection{Selection criteria}
Having reviewed briefly the ``case for in-vacuo dispersion for neutrinos",
and the characterization of its statistical significance provided in  our Ref.~\cite{Ryan},
we are ready to proceed with our analysis of
the ``case for in-vacuo dispersion for photons", emerging from  the investigations reported in
Refs.~\cite{MaZhang,MaXuPRIMO,MaXuSECONDO}. For this photon case, while Ref.~\cite{MaXuSECONDO} convincingly
characterized the relevant feature as striking, there was so far no
characterization of the statistical significance, so one of our main objectives here is
to provide such a characterization.

The analyses reported by Ma and collaborators in Refs.~\cite{MaZhang,MaXuPRIMO,MaXuSECONDO}
focus on the highest-energy photons among those observed for GRBs by the Fermi telescope,
and implements some time-window selection criteria.
Evidently, in spite of the many differences between the two contexts,
there are challenges in this sort of analysis of GRB photons, which are rather similar to
the challenges faced in the analysis of candidate GRB neutrinos reported above.

We find appropriate to here contemplate not only the energy-window and time-window criteria adopted
by Ma and collaborators but also to propose some alternative criteria of our own, which (while keeping close
to the criteria introduced by Ma and collaborators) we feel might be a natural alternative to be considered
as this research program further advances, especially as new data are accumulated.

Ma and collaborators focus on GRB photons
observed within 90
seconds of the first peak in the GBM
and with observed energy greater than 10 GeV. In our alternative criteria we choose to specify the time window
by mainly exploiting the fact that, as already observed in Ref.~\cite{MaXuSECONDO}, a surprisingly high percentage of
the photons selected by the criteria of Ma and collaborators are consistent with roughly the same
value of the
time offset at the source $t_{off}$. We attempt to exploit this aspect in our time-window
selection criteria by essentially characterizing the time window in terms of emission times,
rather than observed times.
We require that at the source the time of emission of our selected photons be consistent with
an offset with respect to the time of emission of the first GBM peak of up to 20 seconds, but of course
also allowing in addition for a sizeable range of effects possibly due to in-vacuo dispersion.
When expressed in terms of the difference $\Delta t$ between the time of observation of the relevant photon
and the time of observation of the first GBM  peak, our time selection criterion takes the
form
\begin{equation}
|\Delta t| \leq 10^{-16} D(z) + (1+z) 20 s \, .
\label{timeselect}
\end{equation}
Here the $20s$ are our mentioned window on $t_{off}$, while the parameter we fix at $10^{-16}$
allows for in-vacuo-dispersion effects of amount roughly comparable to the corresponding range
of effects probed by Ma and collaborators. The main difference here is that our time window
has the same quantitative interpretation for all GRBs when described in terms of emission times at the source,
but when expressed as a window on observed times it depends on the redshift of the GRB. The 90 seconds
of  redshift-independent
observed-time window adopted by Ma and collaborators roughly coincide with our window on observed times
at redshift of 1. For GRBs at redshift greater
than 1 (where both time
dilatation of the offset and the possible in-vacuo dispersion would produce bigger effects
on the time of observation) our Eq.(\ref{timeselect}) allows for an
observed-time window larger than 90 seconds,
while for GRBs at redshift smaller
than 1 it allows for an
observed-time window smaller than 90 seconds.

For what concerns our window on photon energies, consistently with our focus on properties at the source
(rather than observed properties), we require that our selected photons be emitted at the source
with energy greater than\footnote{For what concerns this energy-selection criteria,
we should mention that as we were finalizing the study here reported, in private conversations
with Bo-Qiang Ma, we learned that Ma and collaborators are independently contemplating the
possibility of implementing the selection in  terms of energy at emission, also leaning toward
the possibility of setting the cut at 40 GeV.} 40 GeV. So in terms of observed energy our window is $E \geq 40\text{GeV}/(1+z)$,
an alternative to the 10-GeV redshift-independent observed-energy window of Ma and collaborators.
We picked 40 GeV as our cut on the energy at the source because the selection process
for this choice gives results rather close to those obtained with the cut at 10 GeV of observed
energy adopted by Ma and collaborators.

At the present time (as confirmed by our analysis) there is no evidence that our criteria might be
more advantageous than those of Ma and collaborators. We are only proposing them as an alternative which
might play a role as this research program advances. Accordingly, while we keep at center stage our proposed
criteria, in this manuscript we shall also report some results that we obtained using the selection criteria
of Ma and  collaborators.

An important final remark on selection criteria concerns redshifts. For our neutrino analysis it was
possible, as shown above, to allow for GRB-neutrino candidates for which the GRB redshift had not been measured.
We expect, as argued more extensively in Ref.\cite{Ryan}, that by using as reference some
estimated average value of redshifts for long and short GRBs observed in neutrinos we should eventually,
as more data is accrued, reach conclusive findings,  in  spite of handling GRBs which, for the most part,
have no precise redshift assignment. Such conclusive findings would have been reached faster in presence
of more measured values of redshift, but without such measured values the analysis still works in the long run.
We believe, however, that for
the analogous photon analyses  the role of redshift measurements must be
handled differently. A challenge for this sort of photon analyses is that the size of the conjectured effects,
often of a few seconds, is comparable to the time scales of the astrophysical mechanisms at work
in a GRB. Any eventual in-vacuo dispersion effect would  have to be deduced finely within the
sort of ``background noise" produced by the (largely unknown) mechanisms that cause the specific
time variability of a given GRB. As a result we propose that in-vacuo-dispersion photon analyses
should confine themselves to GRBs of measured redshift. Also Ma and collaborators
rather strictly adopt this attitude
toward redshifts, though they have handled cases\footnote{The only case of this type included so far
in studies
by Ma and collaborators is GRB140619b~\cite{MaXuPRIMO,MaXuSECONDO},
a GRB for which no redshift measurement is available. We shall here not consider GRB140619b.}
where the GRB redshift had been guessed on the basis of some theoretical argument but had not been
measured. We shall assume that it is safer for photon analyses to  focus strictly
on GRBs on measured redshift.

\subsection{Properties of selected photons and statistical analysis}
We show in table 2 and figure 2 the 11 Fermi-telescope photons selected by the time
window of our Eq.(\ref{timeselect}) and our requirement of an energy of at least 40 GeV at emission.
The fact that our criteria are to a large extent compatible with the
criteria of Ma and collaborators is also suggested by Figure 2: all our 11 photons were also
selected by Ma and collaborators; the only difference is that 2 of the photons
selected by Ma and collaborators
are not picked up by our criteria. These 2 additional photons are also shown in Figure 2 and Table 2.

\begin{table}[htbp]
\centering
{\def\arraystretch{0.5}\tabcolsep=3pt
\begin{tabular}{|c|c|c|c|c|c|c|}
\hline
    & $E_{\text{em}}$[GeV]  & $E_{\text{obs}}$[GeV] & $E^*$[GeV] & $ \Delta t $ [s] & $z$ & GRB \\ \hline
1   & 40.1  & 14.2 & 25.4  &  4.40 & 1.82 & 090902B \\ \hline
2   & 43.5  & 15.4 & 27.6  & 35.84 & 1.82 & 090902B \\ \hline
3   & 51.1  & 18.1 & 32.4  & 16.40 & 1.82 & 090902B \\ \hline
4   & 56.9  & 29.9 & 26.9  & 0.86  & 0.90 & 090510  \\ \hline
5   & 60.5  & 19.5 & 40.0   & 20.51 & 2.11 & 090926A \\ \hline
6   & 66.5  & 12.4 & 47.1  & 10.56 & 4.35 & 080916C \\ \hline
7   & 70.6  & 29.8 & 40.7  & 33.08 & 1.37 & 100414A \\ \hline
8   & 103.3 & 77.1 & 25.2 &18.10 & 0.34 & 130427A \\ \hline
9   & 112.5 & 39.9 & 71.5  & 71.98 & 1.82 & 090902B \\ \hline
10  & 112.6 & 51.9 & 60.7 & 62.59 & 1.17 & 160509A \\ \hline
11  & 146.7 & 27.4 & 104.1  & 34.53  & 4.35 & 080916C \\ \hline
12* & 33.6  & 11.9 & 21.3  & 1.90   & 1.82 & 090902B \\ \hline
13* & 35.8  & 12.7 &  22.8  & 32.61 & 1.82 & 090902B \\ \hline
\end{tabular}
}
\caption{Here reported are some properties of the 13 photons picked up by the selection
criteria of Ma and collaborators. Our selection criteria pick up 11 of these 13 photons
 (we place an asterisk on  the 12th and 13th entries in the table in order to highlight that
they are  not picked up by our selection criteria). The second and third columns report respectively the values of
energy at emission and energy at observation. The fourth column reports the difference in times of observation
between the relevant photon and the peak of the GBM signal. The last column identifies the relevant GRB,
while the fifth column reports its redshift.}
\label{table1}
\end{table}

The content of figure 2, as already efficaciously stressed in Ref.~\cite{MaXuSECONDO}, is rather striking.
Following Ma and collaborators, we notice that all 13 photons (the 11 picked up by our criteria, plus the
two additional ones picked up by the criteria of Ma and collaborators) are well consistent with
the same value of $\eta$, upon allowing for only 3 values of $t_{off}$.
We shall  not however attempt to quantify the statistical significance of this more complex thesis
based on 3 values of $t_{off}$: evidently the
most striking feature is that 8 of our 11 photons (9 of the 13 photons of Ma and collaborators)
are all compatible with the same value of $\eta$ and $t_{off}$. This sets up a rather easy question
that one can investigate statistically: if there is no in-vacuo dispersion, and therefore the correlation
shown by the data is just accidental, how likely it would be for such 11 photons to include 8
that line up so nicely?

\begin{figure}
\includegraphics[scale=0.6]{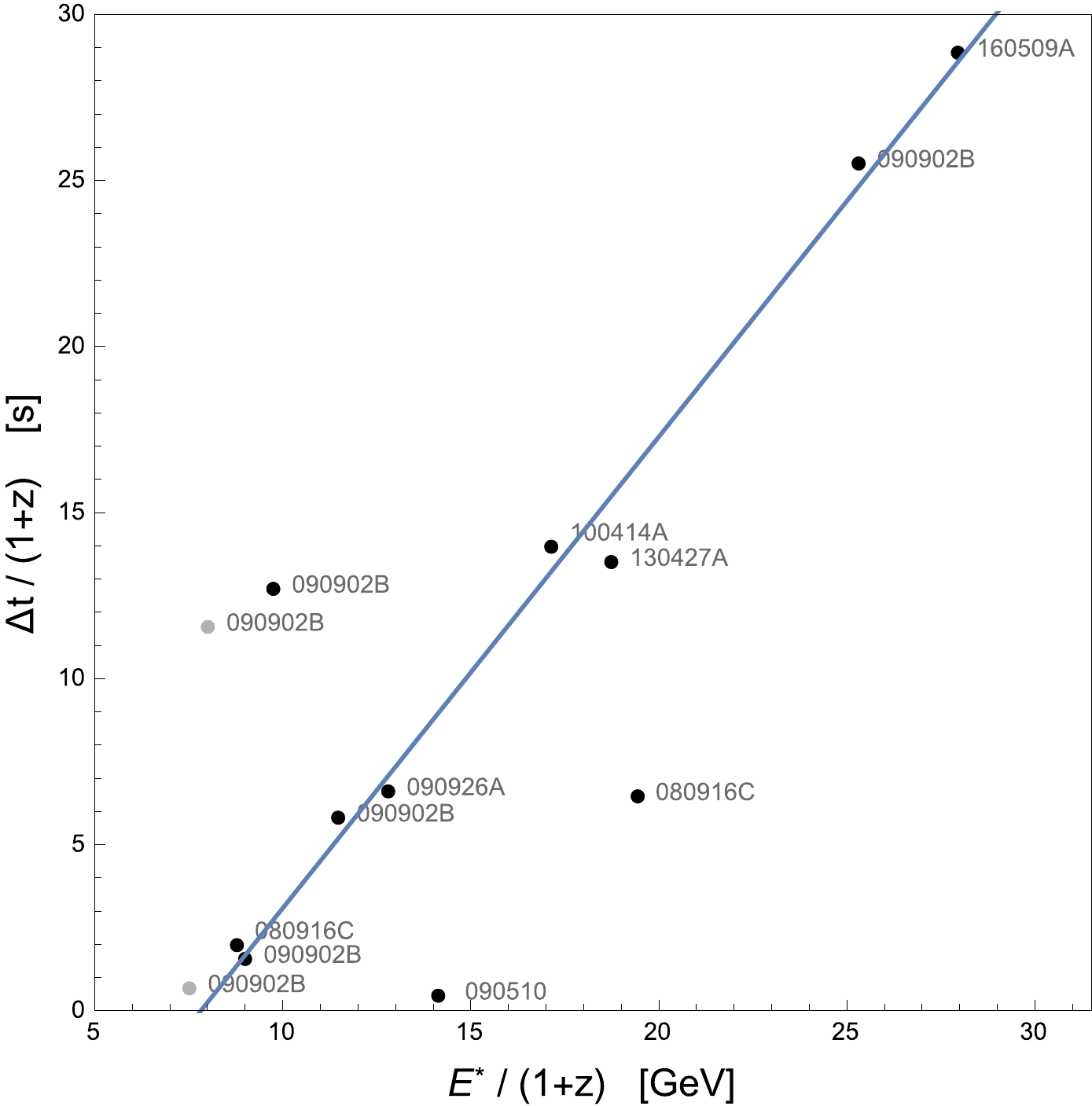}
\caption{Black points here in figure correspond to the 11 photons picked up by our selection criteria,
 characterized in terms of their values of $\Delta t/(1+z)$ and $E^*/(1+z)$.
 Gray points are for the 12th and 13th entries in table 2 (photons picked up by the criteria of Ma and collabroators,
 but not by our criteria).
The strikingly  visible feature of 8 black points (plus 1 gray point) falling nicely on a straight line is also
highlighted in figure by the presence of a best-fit line (which however we find appropriate to discuss
in detail only later, in Section V). }
\end{figure}

We address this question quantitatively by first computing the correlation of the 8 among our 11 photons
that line up nicely in figure 2, finding that this correlation is 0.9959.
We then estimate an associated ``false alarm probability"~\cite{Ryan}
by performing simulations in which (while keeping their energy fixed at the observed value)
we randomize, within the
time window specified by our time-selection criterion,
the time delay of each of our 11 high-energy photons with respect to the GBM peak of the relevant GRB, and
we assign
to each of these randomizations a value of correlation given by the maximum value of correlation
found by taking in all possible ways 8 out of the 11 photons. We find that these simulated values
of correlation are $\geq 0.9959$ only in 0.0013$\%$ of cases, about 1 chance in 100000.

We stress that this impressive quantification of the statistical significance of the feature
exposed by Ma and collaborators does not depend on the fact that we adopted our own
novel selection criteria. For this purpose we redo the analysis including also the 2 photons
that should be included according to the criteria of Ma and collaborators. In this case
we have 9 out of 13 photons that line up very nicely.
The value of correlation found for those 9 photons is  0.9961.
We then perform simulations in which we randomize the time of observation of all the 13 photons within the
time window specified by the time-selection criterion of Ma and collaborators and
to each of these randomizations we assign a value of correlation given by the maximum value of correlation
found by taking in all possible ways 9 out of the 13 photons. We find that these simulated values
of correlation are $\geq 0.9961$ only in 0.0009$\%$  of cases, very close to
the  0.0013$\%$ obtained with our selection criteria.

\subsection{Predictive power}
The values of correlation reported in the previous subsection, and especially the values of false alarm probability
found in the previous subsection, are rather impressive. However, as discussed in the next section,
the interpretation of these data presents us with some challenges.
In light of this we find appropriate to stress that the picture emerging from this photon feature
has intrinsic model-independent ``predictive power". We illustrate this notion
by considering the situation set up by the first two papers by Ma and collaborators, Refs.~\cite{MaZhang,MaXuPRIMO},
which were written before May 2016 ({\it i.e.} before the observation of GRB160509a).
At that point Ma and collaborators had already discussed the photon feature using all the photons
in  our figure 2, of course with the exception of the photon from GRB160509a
which had not yet been observed.
That photon from GRB160509a allowed then Ma and Xu, in Ref.~\cite{MaXuSECONDO},
to appropriately emphasize that the picture was finding additional support.

In a sense which we shall attempt to quantify, the picture Ma and collaborators had been developing
 exhibited some  predictive power upon the observation
of GRB160509a. Our quantification of this predictive power takes off by  computing the value of correlation obtained with the other 8 photons
on the ``main line" of figure 2
({\it i.e.} not including the photon from GRB160509a, but  including the photon on
the ``main line" picked up by the selection criteria of Ma and collaborators but not picked up by our
selection criteria), finding that this correlation is of 0.9935.
With the observation of the photon from GRB160509a the resulting 9-photon correlation moved up to  0.9961.
We shall characterize the predictive power by asking how likely it would be for a photon unrelated to
those previous 8 photons on the ``main line" to produce accidentally such an increase of correlation.
We randomize the time of observation of that photon from GRB160509a (within the
time window specified by the time-selection criterion of Ma and collaborators)
and we find that an increase of correlation from 0.9935 to  0.9961 (or higher)
occurs only in 1.9$\%$ of cases.

We perform the same estimate also
adopting our selection criteria, as a mere academic exercise (our selection
criteria are being proposed here, after the observation of GRB160509a).
With our selection criteria one has only 7 photons on the ``main line", when
 considering data available before GRB160509a. Those 7 photons have correlation of 0.9932.
 Adding the photon from GRB160509a one then has a 8-photon correlation of 0.9960.
We randomize the time of observation of that photon from GRB160509a (within the
time window specified by our time-selection criterion)
and we find that an increase of correlation from 0.9932 to 0.9960 (or higher)
occurs only in 0.79 $\%$ of cases.

\section{Observations relevant for the interpretation of the data}

Our quantification of statistical significance gave rather impressive results both
for the neutrino feature and for the photon feature. We still feel that the overall
situation should be assessed prudently, since both analyses still rely on only a
small group of photons and neutrinos. There is no reason to jump to any conclusions, also
because both the Fermi telescope and the IceCube observatory will continue to report new
data still for some time to come. It is nonetheless interesting to assess the present situation
both from the viewpoint of possible interpretations and from the viewpoint of a possible consistency
between different analyses.

\subsection{Concerning photons outside the ``main line"}
A first step of interpretation must concerns the 3 photons that in  figure 2 do not line up with
the other 8 photons, the 8 photons which lie on the ``main line"~\cite{MaZhang,MaXuPRIMO,MaXuSECONDO}
of Ma and collaborators. The tentative interpretation one must give within the setup of these analyses
is that those 3 photons were not emitted in coincidence with the fist peak of the GBM signal.
The time window of our selection criterion (and similarly the one of the selection criterion adopted
by Ma and collaborators) is structured in such a way to  ``catch" those high-energy photons that were emitted
roughly at the same time when the first peak of the GBM was emitted, but if truly in-vacuo dispersion is at work
evidently it would happen occasionally that just because of in-vacuo dispersion some photons not emitted
in coincidence with the first peak of the GBM end up being observed within our time window. While it is evidently
difficult to quantify how frequently this should occur, at least qualitatively what is shown in figure 2 is just
what one should expect if in-vacuo dispersion truly occurs, including the presence of some photons
outside the ``main line".

As an aside, let us however notice that the significance of what is shown in figure 2 is not washed away
if we include in the analysis of statistical significance also the photons outside the ``main line".
For this purpose we first notice that the value  of correlation obtained by taking into account
all 11 photons is 0.845, still rather high.
In our simulations, in which
we randomize the time of observation of the 11 photons (within the
time window specified by our time-selection criterion),
we find that a value of correlation for all 11 photons $\geq$  0.845
is obtained only in 0.035 $\%$ of cases.

Similar conclusions are reached adopting the criteria of Ma and collaborators. In that case one
has 13 photons under consideration, and the value of correlation computed for those 13 photons is 0.805.
Randomizing the times of observation of those 13 photons (within the
time window specified by the time-selection criterion of Ma and collaborators)
we find that a value of correlation for all 13 photons $\geq$ 0.805
is obtained only in 0.037$\%$ of cases.

\subsection{Trigger time without offsset}
In  light of the  observations made in the previous subsection one feels encouraged
to set aside the 3 photons
that fall off the ``main line", and focus on the other 8 photons.
A significant characterization of those 8 photons
is obtained by assuming $\delta_\gamma =0$, so that the whole feature is due to a nonzero value
for $\eta_\gamma$. This assumption $\delta_\gamma =0$ is very restrictive but still the ``main line"
of 8 photons in figure 1 is very well described by the model of Eq.(\ref{mainnew}), for $t_{off}= - 11 s \pm 1 s$
and $\eta_\gamma = 34 \pm 1$. These are the parameters of the line shown in figure 2, where the goodness
of the fit of the 8 photons on the ``main line" is visible.

The story with the ``main line", the single time offset shared by 8 photons, and $\delta_\gamma =0$ fits indeed in remarkably nice way,
in spite of being based on very restrictive assumptions.
This is surely striking,
but we are nonetheless inclined to proceed cautiously. There is evidently a pronounced feature, of the type here characterized,
in these available GRB-photon data, but its description does not necessarily have to be the one
that presently fits the data so nicely. First we should
stress that in spite of our impressive findings for the false alarm probabilities,
we still consider as most likely the hypothesis that the feature is accidental, rather than a truly
physical (in-vacuo-dispersion-like) feature. Perhaps more surprisingly, even when taking as temporary
working assumption
 that the feature is physical
 we give priority to the hypothesis  that the feature might not
really be describable in terms of the ingredients composed by the ``main line",
the same time offset for 8 photons
and $\delta_\gamma =0$. We are inclined to adopt this attitude because our intuitive assessment
is that, even if the overall feature is physical, at least part of present picture, with these few
data available, could be accidental. We feel that such level of prudence is methodologically correct
in general, and in this case might find additional motivation in the
fact that the offset time favored by the analysis summarized in figure 2 would require,
as observed above, a majority of our photons to have been emitted at the source some 11 seconds before
the time of emission of the GBM peak. (We might have had a slightly different intuition had we found
a similar result but for 11 seconds after the GBM peak.)

We give tangibility to these considerations by taking temporarily as working assumption,
as an illustrative example,
a hypothesis such that the feature is truly physical but the way it manifested itself so far
is in  part accidental. For this purpose we  ``scramble" the nice picture of
figure 2 by not taking under consideration the $\Delta t$, time difference with respect to time of observation
of GBM peak, but rather a $\Delta t_{trigger}$, time difference with respect to trigger time of the
 GBM signal. We do this just to probe the dependence of our results on the perspective adopted in the analysis:
 we would not really expect that $\Delta t_{trigger}$ is better than $\Delta t$
 at exposing the sought correlation, but it is interesting for us to see whether the feature
 completely disappears by replacing $\Delta t$
 with $\Delta t_{trigger}$.

\begin{figure}
\includegraphics[scale=0.6]{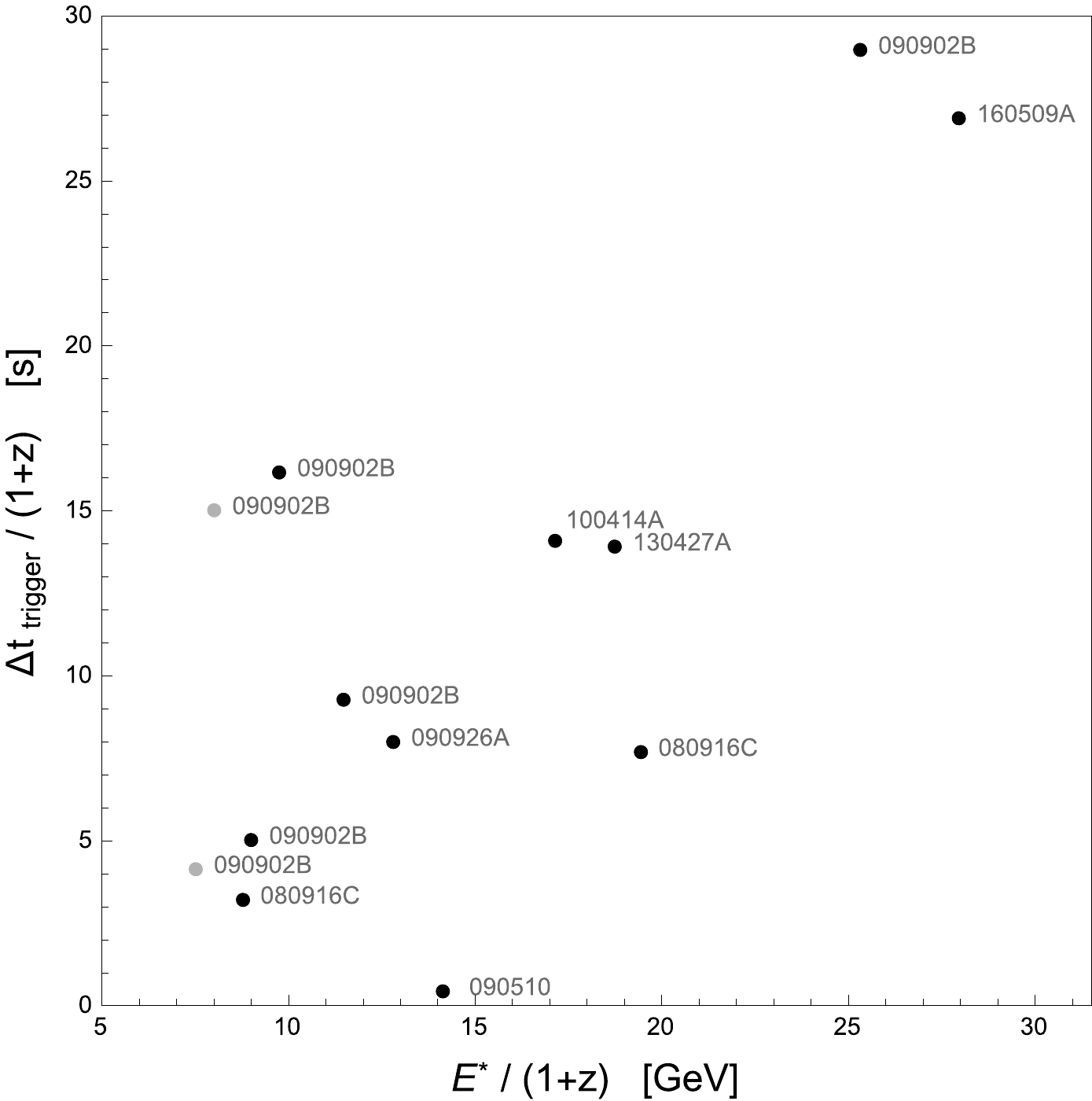}
\caption{Same as figure 2, but replacing $\Delta t$ with $\Delta t_{trigger}$.}
\end{figure}

 What we get upon relying on  $\Delta t_{trigger}$ in place of $\Delta t$
 is the picture given by figure 3.
In figure 3 there is no neat ``main line", but this is after all what we would have expected
before looking at the data: we would have expected the time offset at the source
(with respect to the first GBM peak or to the GBM trigger)
to be at least a bit different for different photons; moreover, with a nonvanishing $\delta_\gamma$ the time of observation
of each photon would receive an additional random component.

Importantly for our purposes, one should notice that, while figure 3 is surely less striking than figure 2,
the feature has not disappeared:
it is less pronounced but the overall picture of figure 3 still shows a surprisingly
high correlation. This is what we mean by contemplating the hypothesis that only part of what is shown
in figure 2 might be physical, with the rest being just accidental result of how these first 11 photons
usable for our purposes happened to match very neatly a particular set of hypothesis for the interpretation
and the analysis.
Quantitatively we have that for
 the data analyzed in the way reflected by figure 3 we have correlation
of 0.775, over all 11 photons picked up by our selection criteria.
Randomizing, within the
time window specified by our time-selection criterion,
the time delay of each of our 11 high-energy photons with respect to the GBM trigger of the relevant GRB,
we find correlation $\geq 0.775$
in only $0.17 \%$ of cases, which is (not as small as the $0.0013 \%$ found above for the analysis
with $\Delta t$, but)
still a very small false alarm probability.

\subsection{Consistency between the features for photons and the feature for neutrinos}
In  light of the  observations made in the previous subsection we now set aside the 3 photons
that fall off the ``main line", and focus on the other 8 photons.
A significant characterization of those 8 photons
is obtained by assuming $\delta_\gamma =0$, so that the whole feature is due to a nonzero value
for $\eta_\gamma$. This assumption $\delta_\gamma =0$ is restrictive but still the ``main line"
of 8 photons in figure 1 is very well described by the model of Eq.(\ref{mainnew}), for $t_{off}= -11 s \pm 1 s$
and $\eta_\gamma = 34 \pm 1$.

It is interesting to compare this estimate of $\eta_\gamma$ with the estimate of $\eta_\nu$ that one can
obtain from the neutrino data here discussed in Section III. This comparison should be handled with some care,
since some quantum-spacetime models predict (see, {\it e.g.}, Ref.~\cite{gacLRR} and references therein)
independent in-vacuo dispersion parameters for different particles, and also a possible dependence of the
effects on polarization for photons and on  helicity for  neutrinos. Still one would tentatively expect
comparable magnitude of the effects for different particles (including the possible dependence on polarization/helicity).
A first important observation is the figure 1 includes~\cite{Ryan} 5 neutrinos whose interpretation in terms of
in-vacuo dispersion would require positive $\eta_\nu$ and 4 neutrinos
whose interpretation in terms of
in-vacuo dispersion would require negative $\eta_\nu$ (this is why in figure 1 we consider~\cite{Ryan}
the absolute value of $\Delta t$). Another complication for our purposes originates in the fact that,
as mentioned,
we have reasons~\cite{RyanLensing} to expect that 3 or 4 of those 9 GRB-neutrino candidates
are actually background neutrinos that happened to fit accidentally our profile of a GRB-neutrino candidate.
What we can do is to attempt an estimate of the absolute value $|\eta_\nu|$ and to perform this estimate
by assuming that 3 of the 9 GRB-neutrino candidates are background: essentially we estimate $|\eta_\nu|$
for each possible group of 6 neutrinos among our 9 GRB-neutrino candidates, and we  combine these estimates
into a single overall estimate. This leads to the estimate $|\eta_\nu| = 19 \pm 4$.

So we have an estimate of $\eta_\gamma = 34 \pm 1$ and an estimate of $|\eta_\nu| = 19 \pm 4$, which
are closely comparable, as theoretical prejudice would lead us to expect. Perhaps more importantly, the hypothesis
that both features are accidental should also face the challenge introduced by this correspondence of
values. If actually there is no in-vacuo dispersion both features should be just accidental.
All 9 of our GRB-neutrino candidates would just be background neutrinos who happened to fit our
criteria for selection of GRB-neutrino candidates and whose energies and times of observation just happened
to produce the high correlation shown in figure 1. And similarly all 11 of the photons selected by  our criteria
would have accidentally produced the correlation visible in figure 2: they would be photons whose time of observation
(with respect to the time  of observation of the GBM peak) is not really correlated with energy, the correlation
with energy emerging just accidentally. All these assumptions about neutrinos and photons are needed
if there is no in-vacuo dispersion, with the additional observation that all these accidental
facts end up producing comparable estimates of $\eta_\gamma$ and $|\eta_\nu|$.

The level of ``consistency" (in the sense discussed above)
between the neutrino feature and the photon feature is visually illustrated in our figure 4, showing both
our 11 photons and our 9 GRB-neutrino candidates in a plot of $E^*$ versus the absolute value
of $\Delta t$.

\begin{figure}
\includegraphics[scale=0.6]{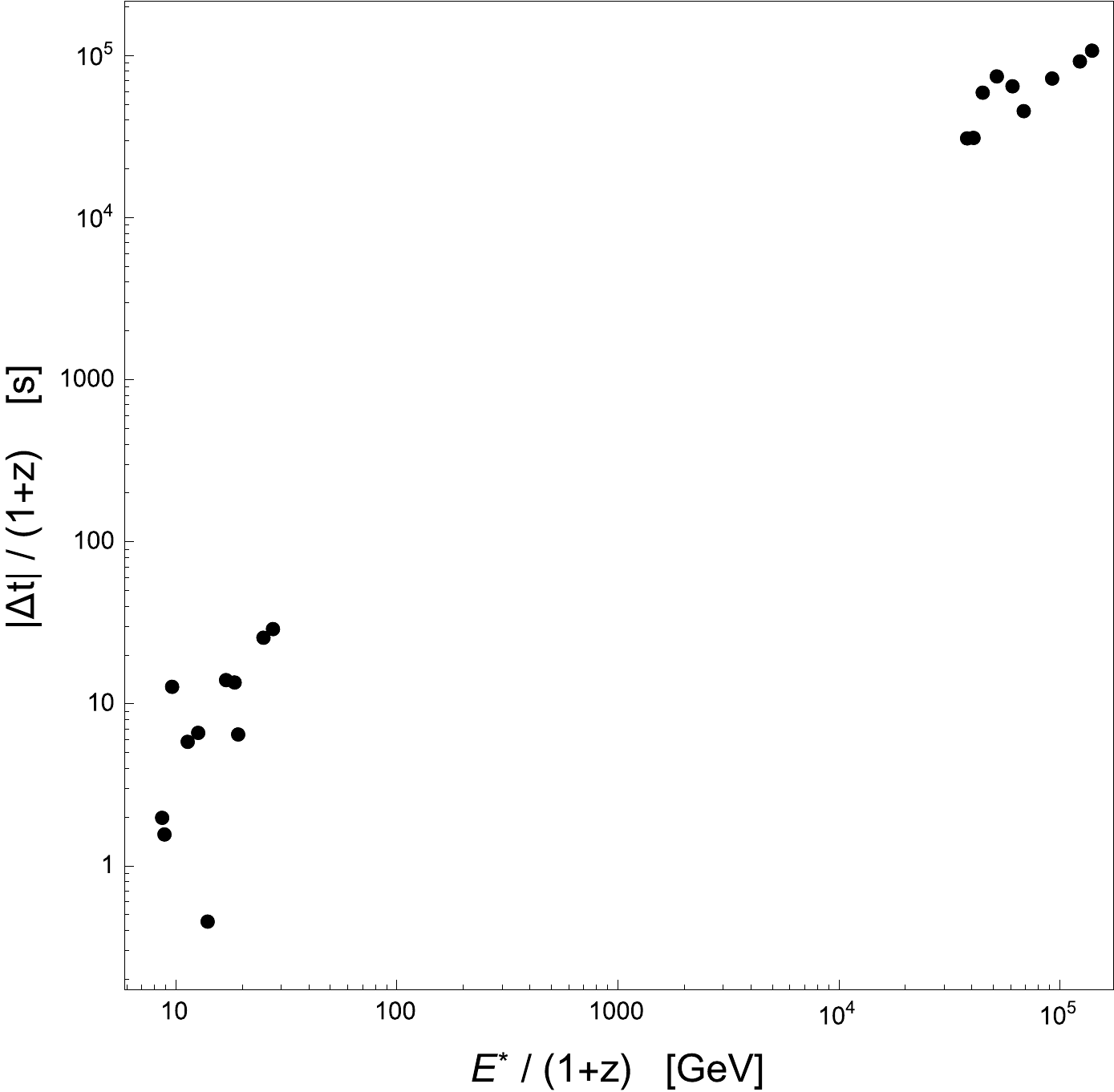}
\caption{Here we show together the content of figures 1 and 2, so that one can appreciate
the overall picture for what concerns the
correlation between $|\Delta t|$ and $E^*$.}
\end{figure}

\subsection{On a possible astrophysical interpretation of the photon feature}
So far we only considered two alternative hypotheses: either the two features shown in figures 1 and 2 are due to in-vacuo
dispersion or there is no in-vacuo dispersion and those two features are accidental.
One should of course contemplate a third possibility: there might be no in-vacuo dispersion but still at least one of
those two features is not accidental, but rather the result of some other physical mechanism.
This leads us naturally to wonder whether the two features could be the result of some (so far unknown)
 astrophysical properties of the sources.

We believe the hypothesis that the neutrino feature be of astrophysical origin should be discarded:
the relevant effects are of the order of a couple of days, and neutrinos observed two days before or after
a GRB could not  possibly be GRB neutrinos (unless in-vacuo dispersion takes place).
If the neutrino feature is  confirmed when more abundant data become available we would know that
something not of astrophysical origin has been discovered.

In this respect the photon feature is very different. The size of the effects is between a few and $\sim 100$ seconds,
which may well be the time scale of some mechanisms intrinsic of GRBs. The main reason to be skeptical
about the astrophysical interpretation comes from the fact that the content of figure 2 reflects the properties
of the $D(z)$:  the data points (those on the ``main line") line up  only because we have factored the $D(z)$
in  the analysis, and the $D(z)$ is a form of dependence on redshift which reflects propagation. So the astrophysical
interpretation of the photon feature still requires assuming that at least part of the content
of figure 2 is accidental: we cannot exclude some mechanism at the GRB producing some level of correlation
between energy of the photon and difference in time of emission with respect to the GBM peak, but
such a mechanism could produce the feature of figure 2 only if accidentally (on those few data points)
it ended up taking values lending themselves to the sort of $D(z)$-dependent analysis
which we performed.

So the astrophysical interpretation of the photon feature is possible but must face some issues.
However, we should stress that also the interpretation of the photon feature in terms of
the model (\ref{mainnew}) has to face a challenge connected with GRB090510. One of 3 photons
off the ``main  line" is a 30 GeV photon from the short GRB090510. As discussed above, when taking
as working assumption that
in-vacuo dispersion actually takes place, these photons off the ``main  line" should be interpreted as photons
emitted not in (however rough) coincidence with the first peak of Fermi's GBM. Such an interpretation
is certainly plausible in general, but the case of the 30 GeV photon from GRB090510 is a challenge.
That 30GeV photon was observed~\cite{fermiNATURE} within the half-second
time window where most GRB090510 photons with energy between 1 and 10 GeV were also observed.
In light of this, it is certainly very natural to assume that the 30GeV photon could not have accrued an
in-vacuo-dispersion effect of more than half a second, travelling from redshift of 0.9 (the redshift of GRB090510),
which implies $|\eta_\gamma| <1$. For $\eta_\gamma \sim 30$, as suggested by the  points on the ``main  line",
the in-vacuo-dispersion effect for that 30GeV photon should be of more than 15 seconds. It should have arrived together
with that half-second-wide peak of 1-10GeV photons because of an accidental and strong cancellation
between an effect of $\sim 15$ seconds due to emission-time differences at the source and a 15-second in-vacuo-dispersion effect accrued propagating.
This is certainly possible, but a bit ``too lucky" for our taste.

We feel the
the 30 GeV photon from GRB090510 poses a very severe challenge for the interpretation of the photon
feature in terms of our model (\ref{mainnew}), even though all other photons in our data
fit so nicely (\ref{mainnew}). In connection with this one should notice that the 30GeV photon is
the  only photon in our sample coming from a short GRB (GRB090510). All other photons in our sample come from
long GRBs.
If the effect is present for long GRBs and absent for short GRBs, then the interpretation should
be astrophysical. One can also notice however that GRB090510, with its redshift of 0.9, is one of the closest
GRBs relevant for our photon analysis.  All other GRBs in our photon analysis, with the exception of GRB130427a,
are at redshift greater than 1.  A scenario in which the effect is pronounced only at large redshifts
could be of quantum-spacetime origin, but of course would require a quantum-spacetime picture in which
the dependence on redshift of the effects is not exactly  governed by the function $D(z)$.

\section{Closing remarks}
More data will soon be available both for our photons and for our neutrinos,
so we shall not dwell much on the significance of our findings.
We just stress that surely the false alarm probabilities here derived are small enough to
motivate further interest in this type of analyses.
Particularly for neutrinos a much improved analysis should become soon possible,
since so far IceCube only made publicly available their data up to May of 2014,
so at the time of writing this article we know that some additional 2.5 years of data
have been collected by IceCube but have not yet been publicly released.
For photons our main reference is the Fermi telescope, which has been operating
since 2008. In about 8 year of operation Fermi provided 7 GRBs contributing
to the photon side of our analysis (see table 2), so we can expect to have  roughly
one GRB per year adding points to our figure 2.

As stressed above, if the neutrino feature was confirmed it would be very hard to
even imagine an astrophysical origin for that feature. For photons instead our intuition,
while being open to ultimately finding conclusive evidence of in-vacuo dispersion,
presently favors the possibility of a scenario in which the feature is confirmed by additional data but
in  the end the correct description be
given in terms of some properties
of the astrophysical sources. We would welcome feedback from the astrophysics community
on the type of ``mechanisms at the source" that could produce such a feature for photons.
On the other hand overall, combining both the neutrino side and the photon side of our analysis,
it turn out that the feature is stronger at higher energies and higher redshifts,
so
we feel that our findings could motivate the development by the quantum-gravity community
of models similar to the one of Eq.(\ref{main})
but such that the effects are indeed less pronounced than predicted by
Eq.(\ref{main}) at lower energies and/or lower redshifts.

\section*{Acknowledgements}
We are very grateful to Bo-Qiang Ma and Simonetta Puccetti for valuable discussions on some of data here used.
 We also gratefully acknowledge conversations with Fabrizio Fiore
and Lee Smolin.
The work of GR  was supported  by funds provided by the National Science Center under the
agreement DEC- 2011/02/A/ST2/00294.
NL acknowledges support by the European Union Seventh Framework Programme (FP7 2007-2013) under grant agreement 291823 Marie Curie FP7-PEOPLE-2011-COFUND (The new International Fellowship Mobility Programme for Experienced Researchers in Croatia - NEWFELPRO), and also partial support from the H2020 Twinning project n$^o$692194, "RBI-TWINNING".

\end{document}